# Explanation Hacking: The perils of algorithmic recourse

Emily Sullivan
Utrecht University

Atoosa Kasirzadeh
University of Edinburgh

**Abstract:**
We argue that the trend toward providing users with feasible and actionable explanations of AI decisions—known as recourse explanations—comes with ethical downsides. Specifically, we argue that recourse explanations face several conceptual pitfalls and can lead to problematic explanation hacking, which undermines their ethical status. As an alternative, we advocate that explanations of AI decisions should aim at understanding.

## 1. Introduction

When an AI system recommends against you getting a job interview, a bank loan, or categorizes you as 'high risk,' you probably want to know *why*. At first glance, it might seem simple to give reasons why, but in reality, there are several different kinds of reasons, including a myriad of justifications, that could fulfill this explanatory purpose. These *reasons-why* serve as explanations. In the context of AI, particularly opaque algorithms, explainable AI techniques are employed to articulate such reasons behind the AI's outputs. One primary research directive on the explainability of AI systems focuses on questions concerning the (non-)epistemic norms required to satisfy our desire–and right–to know why an AI system made its decision. One normative approach that is gaining more and more traction is *algorithmic recourse* (Ustun et al., 2019; Venkatasubramanian and Alfano, 2020). Recourse explanations give reasons that are actionable and feasible for the end user or data subject to change. If the end user takes on these actionable reasons by implementing changes in their life, then, as the story goes, they would get a decision reversal in the future from the same model, *ceteris paribus*. Take for example being rejected for a bank loan. A recourse explanation of this decision could be something like the following: If you had worked full-time, instead of part-time, then you would have been approved for the loan. This explanation suggests that the user could transition to full-time work and secure a loan approval upon reapplication.

Algorithmic recourse has been held up as an ethical gold standard for AI explanations. For example, Venkatasubramanian and Alfano (2020) say the following:

> Recourse systematically delivers the benefit of reversing harmful decisions by algorithms and bureaucracies across a range of counterfactual scenarios. If someone enjoys recourse, then not only are they able to get a single decision reversed, but they also enjoy the power to reverse decisions across a range of counterfactual scenarios. As such, someone who enjoys recourse need not passively suffer the slings and arrows of outrageous fortune, but is positioned to take up arms against a sea of troubles.

Let's call this sentiment the recourse-first view for explainable AI.

**Recourse-first:**
Providing users with recourse explanations of AI decisions—i.e. feasible and actionable counterfactual explanations—is (ethically) preferable to providing users with non-recourse explanations.

In this chapter, we argue that the recourse-first view should be rejected. Instead, we advocate for the following understanding-first view:

**Understanding-first:**
Providing users with explanations that aim at understanding AI decisions is (ethically) preferable to providing users with recourse explanations.

While providing users with recourse explanations seems like the ethically responsible strategy, we suggest that, in reality, recourse explanations face several pitfalls and hidden assumptions that undermine their ethical status. An understanding-first approach, we argue, has a better chance of fulfilling our need, and right, to know why an AI system made the decision it did.

A few points of clarification are in order. First, in this chapter, we are concerned with explaining AI systems behavior (what we can call *model-model* questions). We are not discussing how we might use an AI system to explain some real-world phenomena (what we can call *model-world* questions).[1] Second, in this chapter, we focus on counterfactual explanation methods in xAI. We do this because the work on recourse emerged out of counterfactual explanation methods. That said, we think our arguments also generalize to other methods (see section 5.2).

The chapter proceeds as follows. First, we introduce counterfactual explanation methods and why they have become a popular explanation method for AI decisions. We discuss that counterfactual explanations can have several aims that come apart (section 2). Second, we discuss the recourse-first view, why it seems attractive, but also argue that it faces many problems (section 3). Third, we turn to the understanding-first view. We describe the view and how this view relieves some of the problems of recourse (section 4). Lastly, we discuss possible objections, open questions, and directions for future work (section 5).

## 2. Counterfactual explanations

Counterfactual explanation (CE) methods are a type of post-hoc explainability method that seeks to give insight into the input features influencing AI systems' decisions (Biran and Cotton, 2017; Karimi et al. 2020; Wachter et al. 2017). Counterfactual explanations seek to answer "what if" questions by pointing to minimal changes in the input features that flip the prediction. As a result, CE methods do not provide the direct reason why AI systems make decisions but provide modal information to understand the decision boundary. For example, consider a bank loan AI decision support tool. For any decision, CE methods will perturb various input features and combinations of features to see what minimal changes will turn a loan rejection into a positive decision. The resulting CE might look something like: "If you carried $10,000 less debt and had a salary of over $50,000, you would have gotten the loan." The idea is that having this modal information can help users understand why the system made the decisions it did.

CE methods boast that they don't have the same issues of model fidelity that other post-hoc explanation methods have since they are perturbing on the input variables of the learned ML model directly (Mothial et al. 2020, Mittelstadt et al., 2019). However, CE methods can still engage in distortions. They must select which counterfactual scenarios are the most salient from a larger set of possible counterfactuals. Many CE methods assume causal independence of input features, thus can provide explanations that are impossible in the real-world (Hooker et al. 2021, Karimi et al. 2020). For example, a system suggesting, in a medical context, to increase one's weight but lower BMI. However, weight and BMI are entangled, making it impossible to do both. CE methods suffer from the *Rashomon effect* where different counterfactuals explaining the same decision give conflicting and mutually exclusive reasons (Molnar 2020). Some recent work even suggests that CEs can be manipulated (Slack et al. 2021). However, in this chapter, we want to set aside worries about adversarial manipulation and model fidelity. Even if those issues are resolved there are several conceptual issues and pitfalls that face CE methods, especially recourse CEs.

---

[1] For a discussion on how xAI methods may fail to answer model-world questions, see Durán (2021).

Since CE methods probe the AI system for possible decision flips, CE methods can generate *many many* different counterfactuals of the same decision. Depending on how complex the ML model is, there could be thousands or more different counterfactuals for a single decision. Thus, one of the central challenges of CE methods is filtering these counterfactuals so that they meet particular high-level goals. In the landmark paper on CE methods, Wachter et al. (2017) highlight three distinct aims for CEs. i) elucidating why a decision was reached; ii) offering grounds to contest the decision; iii) suggesting actionable changes to reverse the decision. While Wachter et al. (2017) suggest that any given CE could satisfy all three of these aims, we think it is best to think of these aims separately. Table 1 breaks down these three top level aims and identifies their underlying value.[2] In this chapter, we focus on aims i) and iii), showing that they can come apart in interesting ways (König et al. 2023, Freiesleben and König 2023).

**Table 1: Aims of counterfactual explanations**

| Top level aim | Underlying value |
|---|---|
| Why a decision was reached | Understanding |
| Providing grounds to contest the decision | Informed self-advocacy |
| Providing actionable changes to reverse the decision | Algorithmic recourse, Behavioral adaptation guidance feasibility |

Consider again the example of an AI system that makes loan decisions. Imagine that a 65-year-old wishes to take out a loan from a bank and their application gets rejected based on an AI system's output. There are several CEs that would result in decision flips, but to simplify the case, we can just focus on the following two CEs.

**Retirement**: If you were further away from retirement age (i.e. 5-10 years younger), you would have received the loan.

**Debt**: If you had $50,000 less in debt and $20,000 per year higher income, you would have received the loan.

---

[2] There could be even more aims for CE, we leave this discussion for future work (see e.g. Buchholz 2023).

Let's further imagine that in this case, age was the feature that weighed the most for the decision, and that the retirement counterfactual was the closest CE to the ML decision boundary.[3] In this case, the retirement counterfactual, achieves the aim of *understanding* why the decision was reached, but fails to achieve *recourse*. First, the retirement counterfactual fails to achieve recourse because it is not physically possible–despite what some tech billionaires believe–for someone to become 5-10 years younger. However, the retirement counterfactual still provides the loan applicant with understanding the decision because it provides the loan applicant with the most influential feature that determined the loan outcome, namely that they were simply too close to retirement age. Barocas et al. (2020) call explanations that provide users and data-subjects with the crucial or key difference-makers against them a *principal-reasons* based explanation, which is essential for understanding an AI decision (see also Verma et al. 2020).

But what about the debt explanation? First, the debt explanation achieves the goal of recourse, since it could be feasible for the loan applicant to pay off some debt and increase their salary and apply again. However, notice that with the debt CE there is something epistemically suspicious. It fails to provide the loan applicant with the most influential difference-maker for the decision. The principal reasons explanation is absent. Instead, the recourse CE provides (helpful) advice regarding possible ways for reversing the decision. Because of this, Sullivan and Verreault-Julien (2022) have argued that recourse CEs are not actually explanations but should be considered recommendations. Similarly, Karimi et al. (2020) make a distinction between contrastive explanations and consequential recommendations. Contrastive explanations are those that provide information about the relationship between the model and its inputs–thus, aiming to understand the AI decision. And consequential recommendations are recourse CE that requires additional causal information linking up model inputs to the world. It is our contention that the debt CE does not aim at understanding the AI decision because it hides the most important feature further separating the aims of recourse and understanding. Given that CE aims can come apart, one of the central questions for CE methods concerns which aims should CEs methods optimize for. The growing trend in CE research is that the central aim should be recourse.

---

[3] Closeness to the decision boundary is a highly contextual concept that relies on specific similarity metrics, which are themselves contested. There are various ways to define a similarity order (see, for example, Lewis (1979) and Günther (2022)). However, Morreau (2010) demonstrates that there is no single, objective way to compare overall similarity, as similarities cannot be weighed against each other in a universally agreed-upon manner. We leave this issue aside for now and discuss cases even if we had an objective way to measure this closeness we run into problems (see Kasirzadeh and Smart, 2021).

## 3. Recourse Explanations

Algorithmic recourse embodies three different concepts. The first, it identifies which actionable input variables need alteration to modify a prediction (Ustun et al., 2019). The second, it shows the specific actions that the data-subject must undertake to change the prediction (Karimi et al., 2021). The third, it picks on the actions required by the data-subject to change the target variable itself (König et al., 2023). In all these senses, recourse takes on a very specific meaning to provide *feasible and actionable advice that could reverse AI decisions*. This is different from recourse in the sense of *contestability*, where data-subjects might dispute a decision, lodge an objection, or pursue certain legal protections (aim ii) in Table 1). The fact that the CE literature has co-opted the most ordinary connotation of the concept 'recourse' can cause confusion and may even give the appearance of ethical credibility to feasible and actionable CEs more than is merited.

### 3.1. Recourse-first

The recourse-first view for CEs is as follows:

> **Recourse-first:**
> Providing users with recourse explanations of AI decisions—i.e. feasible and actionable counterfactual explanations—is (ethically) preferable to providing users with non-recourse explanations.

There are several reasons why a recourse-first view is attractive. First, recourse can provide data subjects with temporally extended agency. Venkatasubramanian and Alfano (2020) argue via Pettit (2015) that temporally extended agency is a morally robust good. We are creatures that make plans for the future that matter to us. In order to achieve our goals, we often need to make a series of sub-plans that are themselves intertwined. Having the right level of foreknowledge to make reliable judgments regarding which plans can in fact achieve our goals, is a necessary condition for having temporally extended agency. As agents living in complex societies we have a right to be able to adequately know how to plan for things that greatly impact our lives. Recourse, they argue, is ethically valuable, and preferable, because it provides data subjects with knowledge that supports temporally extended agency. Recourse CEs provide information regarding how to change loan decisions, job interview decisions, and more, by giving actionable and feasible advice for decision reversals.

Second, recourse explanations can be easily aligned with data subject's and user preferences. Users could, in theory, tell a CE explainer model which actions they are willing to do to reverse a decision, and the CE explainer can be tailored to these preferences. Aligning with stakeholder preferences is a leading approach for thinking about explainability of AI systems (Langer et al. 2021; Zednik 2021). Recourse CEs seem especially suited toward customizable stakeholder interests. This can in turn increase data subjects' and users' sense of trust and control over the system. If they are given actionable and feasible ways to reverse decisions, then users can find the decisions intuitive, fair, and give them a sense of control.

Third, recourse explanations can be adapted to other ethical theories, such as the capability approach, which can increase the ethical standing of CEs. For example, Sullivan and Verrealut-Julien (2022) argue that in order for recourse CE to live up to their ethical promise, aligning feasibility with the capability approach provides useful ethical standards for recourse.

However, despite the initial appeal of a recourse-first view for explaining AI decisions, recourse CEs face a series of problems that, we argue, should lead us to reject the recourse-first view. In this chapter, we will focus on three central problems: deflecting responsibility, the problem of hidden assumptions, and explanation hacking.[4]

### 3.2. Problem of deflecting responsibility

The first problem with recourse CEs is that they can deflect responsibility away from the AI system or institution that the AI system is a part of, placing responsibility on the end-user or data subject. Recall that the distinctive characteristic of recourse CEs is that they provide data subjects with behavioral adaptation guidance for their life for the purpose of reversing model decisions in the future. In a bank loan case, providing a recourse explanation signals that it is not the bank's responsibility that someone got rejected for a loan, it is the applicant's responsibility to make changes to their financial life to rectify their financial situation, if they want a different decision. While there are indeed cases where the responsibility ought to be placed on end-users or data subjects, there are notable cases where this deflection in responsibility is problematic. To see why, let's first consider a case outside of an AI decision context.

> **Apartment fire:**
> There was a fire in an apartment building, leaving several units uninhabitable. The tenants want an explanation why the fire spread so quickly throughout the building. The building developer and landlord was told by the fire chief that the main reason the fire spread was due to the flammable building materials that went against current fire safety regulations.

It seems straightforward that the counterfactual explanation that provides tenants with understanding why the apartment fire spread as fast as it did would be something like: *If the*

---

[4] There are other problems that can face CE methods, such as fairwashing (Aïvodji et al. 2019) and worries of paternalism.

*building had been built according to the fire safety regulations, then the fire would not have spread*. This CE exposes that the responsibility for the fire damage is on the developer and landlord for violating fire safety standards. If the tenants were given such a CE, then this could trigger certain legal protections and property damage claims. However, notice that this CE does not provide the tenants with any feasible and actionable changes that *they themselves* could have done to prevent the fire. The fire safety regulation CE does not give tenants 'recourse' in the sense discussed in explainable AI. If the landlord was inspired by recourse explanations in AI and wanted to provide a recourse CE to their tenants, it would have to look something like: *If the apartment had not had any furniture or books along the walls, and instead had several fountain walls, then the fire would not have spread*. While strictly speaking true, and feasible for tenants to implement in future apartments (and would have prevented the fire in this apartment), it seems quite absurd that this explanation could ever be appropriate.

The apartment fire case is, in a lot of ways, silly. However, the absurdity of the example highlights what is lurking behind recourse CEs for AI decisions. Recourse CE has an implicit assumption that the underlying AI system is fair, robust, and trustworthy. However, given the well documented problems that AI systems have with bias and robustness, we should not opt for a CE method that assumes otherwise. Looking back to the bank loan case where we are faced with choosing between the retirement CE and the debt CE, the debt CE deflects responsibility away from the bank onto the data subject. However, if the data subject was given the retirement counterfactual, then it exposes values that are embedded within the AI system that might lead to legal or other contestability avenues for the data subject. In the face of uncertainty regarding the fairness and robustness of AI systems, recourse CEs unduly deflect responsibility away from the AI system.

However, even if a model achieves high standards of fairness, robustness, and trustworthiness, there still may be other reasons that a recourse explanation unjustly deflects responsibility. Recourse CE are focused primarily on actions that can be taken by *individuals*; however, sometimes a socio-structural explanation is more appropriate for explaining AI decisions (Smart and Kasirzadeh, 2024). For example, Haslanger (2015) argues that there are cases where a social level explanation is required in order to capture social injustices.[5] AI decisions occur in several domains where there is a history or wide social injustice, such as the COMPAS model that gives risk scores to those who have been arrested for crimes that they will be arrested again in the future (Biddle 2022). Or consider this question: Why are the outputs of widely used AI systems for automated healthcare allocation racially biased? Smart and Kasirzadeh (2024) argue that a socio-structural explanation can reveal the main reasons: These algorithms are embedded in a racially biased economic and social structure that perpetually disadvantages Black individuals and other minorities, leading to poorer health outcomes. This bias is mirrored in the training data of these algorithms through patterns of exclusion, limited access, and inferior care, elements of which are stratified along ethnic and racial lines. In cases where an AI system is deployed in a context rife with social injustice, giving any individual level explanation can be deflecting responsibility away from the true responsibility holder, which in the COMPAS case will likely include neighborhood policing practices.

---

[5]  See Ross (2023) for a discussion of social explanations as a type of causal constraint explanation.

### 3.3. Problem of hidden assumptions

In the previous section, we looked at how recourse CEs, even if they are designed properly, may deflect responsibility away from the true responsibility holder. In this section, we change gears to discuss the problems facing developing recourse CEs in the first place. Specifically, recourse CEs face the serious problem of hidden assumptions. While recourse CEs offer insights into potential decision-altering actions, they often carry implicit assumptions about the feasibility and actionability of these suggested changes. These assumptions can be overly simplistic and may not account for the complex, multifaceted realities of individuals' lives.

The hidden (normative) assumptions underlying counterfactual explanations in explainable AI have not gone unnoticed. Kasirzadeh and Smart (2021) articulate the range of ontological, semantic, and ethical considerations that shape and influence the evaluation of AI CEs in general. For instance, take the manipulation of social categories like gender or sex in a CE. The choice of which alternative worlds to consider for evaluating what would counterfactually happen hinges on a background semantic theory. This theory guides us in determining relevant worlds that align with the social categories in question. However, there might be scenarios where the semantic theory does not support an apt counterfactual intervention involving these categories. Consequently, this disconnect can lead to an inability to properly satisfy the assumptions necessary for evaluating the truth or falsity of counterfactuals.

For example, imagine an AI system analyzing wage disparities. If we counterfactually change a person's gender to assess differences in wages, the validity of this counterfactual depends on whether the chosen alternative world — where the person's gender is different — is deemed relevant and plausible by the semantic theory. If the theory does not recognize gender change as a legitimate variable for this context, the counterfactual intervention fails to provide meaningful insight into the wage disparity. The analysis by Kasirzadeh and Smart's (2021) on the hidden assumptions of CEs is a broader problem beyond recourse, and we will return to this problem in section 4.3. However, their framework for examining the underlying assumptions inherent in CEs is also helpful for isolating the unique problems with recourse CEs.

The first hidden assumption of recourse CE is the potential disregard for systemic and structural constraints that individuals face. Recourse CEs, by design, suggest changes at the individual level. As a result, their content fails to acknowledge or reference societal or economic elements that play a core role in shaping an individual's opportunities. For instance, suggesting a career change via a CE explanation of "If you had a career in engineering instead of teaching, you would have been approved for the loan" ignores systemic issues like employment discrimination, job market saturation, or educational barriers that might impede such a transition. This potential disregard for systemic and structural constraints can place an unfair burden on the individual to overcome systemic barriers that are often beyond their control, but still places the responsibility of the AI decision on that same data subject.

The second hidden assumption concerns the oversimplification of user experiences, often leading to generalized solutions that do not fit all. For instance, recourse CE methods typically use probability distributions to assess the cost of certain actions, but these costs vary greatly among individuals.[6] This variance is due to differences in cultural, socio-economic, and familial backgrounds, making some CE suggestions impractical or irrelevant for specific users. For example, obtaining an additional educational degree might be feasible for someone with ample time and resources, but it is a far more challenging proposition for individuals with limited leisure time or financial constraints. Similarly, when CEs propose lifestyle alterations like financial behavior changes or dietary shifts, they must consider the individual's unique socio-economic circumstances and cultural practices. Suggesting financial savings without understanding a person's income and expenses, or recommending dietary changes without considering cultural eating habits, overlooks the diverse realities in which people operate.

The third hidden assumption pertains to immediate actionability. Often recourse CEs suggest alterations that are assumed to be immediately actionable, disregarding the necessary time, resources, and effort. For example, the suggestion to reduce a significant amount of debt, say \$50,000, presupposes immediate feasibility. However, this overlooks the nuanced individual circumstances and the complexity of factors that contributed to the accumulation of such debt. The assumption that substantial debt reduction can be swiftly actioned fails to recognize the real-world challenges and constraints faced by many users, rendering the suggestion unfeasible and detached from their practical realities.[7]

The fourth hidden assumption concerns normative injection made by recourse CEs. When CEs propose changes that an individual should make, they may (unwittingly) endorse specific norms about financial status, lifestyle choices, or personal decisions. This normative injection can lead to ethical clashes, especially when the AI system's recommendations conflict with an individual's personal values or ethical beliefs.[8] For instance, consider a scenario where a CE suggests that an individual should reduce their debt by cutting down on non-essential expenses. While this advice may seem financially prudent, it inherently carries a judgment about what constitutes non-essential spending. For some, what is deemed non-essential by the system might be culturally or personally significant, such as spending on church donations or family gatherings. In another example, a CE might recommend someone invest more in savings or retirement plans to improve their financial security. This suggestion, while sound in economic terms, overlooks the individual's immediate financial responsibilities or challenges, such as supporting a family or paying off student loans. It also implies a normative judgment about the importance of long-term savings over present needs.

---

[6] Individual characteristics can be integrated into the selection of explanations and metrics, as demonstrated, for instance by (Shang et al., 2022). However, the critical challenge lies in the effective and robust incorporation of these characteristics. While there is ongoing work in this area, conceptual challenges (such as measuring and formalizing all the relevant aspects of user experiences) persist.

[7] Beretta and Cinquini (2023) explore the incorporation of the temporal dimension into causal algorithmic discourse. Despite this, the temporal assumption remains largely implicit.

*3.4. Explanation hacking*

The last problem we consider that faces the recourse-first view is the potential for *explanation hacking*. The problem is as follows. There is an implicit view in work on CE methods and xAI that we need to make sure that AI systems align with stakeholder values. One legitimate way to do this is by tailoring *explanations* to certain values and stakeholder interests. Aligning explanations to stakeholder interests is the received view in xAI research (Langer et al. 2021; Zednik 2021). However, CE methods, as explained above, face a unique challenge here: there are too many CEs that can be generated requiring a method for filtering these counterfactuals. Having to search through thousands of counterfactuals runs the risk of searching the possibility space in a motivated way. Adopting the stakeholder model makes this even more likely. In the stakeholder case, CE methods search for counterfactuals that stakeholders want to find, while discarding those that do not align with their values. One tempting way to address some of the problems with the hidden assumptions of recourse CEs is to make even more tailor-made explanations. While this method might seem like a virtue, it can be easily abused or manipulated.

We want to suggest that CE methods run the risk of engaging in practices that resemble p-hacking in the social sciences (Head et al., 2015). The social sciences in the past 10 years have suffered from false positives and replication failures. In 2014 the term 'p-hacking' was coined by Simonsohn et al. (2014). P-hacking describes the decisions that researchers make that end up intentionally, or not, fishing for significance in a sea of data points. Researchers choosing to stop or continue data collection once they get a result they like, or by changing which variables are being compared, is a way of ensuring a hit on a desired p-value, which is often necessary for publication. But engaging in these practices runs the risk of publishing a false positive, and is seen as a parasitic practice that undermines the overall integrity and credibility of the scientific field as a whole.

---

[8] See Jesse and Jannach (2021) for a review on the way recommendation systems in general can create nudges.

Similarly, generating thousands of counterfactuals for a single decision creates an environment where CE methods are calibrated post-hoc in a motivated fashion. Explainers can fit the method to find the counterfactuals that they are motivated to find that align the interests of modelers, industries, or data-subjects. Companies can look for recourse explanations that hide bias and engage in fairwashing (Aïvodji et al. 2019). Moreover, they can provide different counterfactuals to each stakeholder, and these could be mutually exclusive. While it might seem advantageous that the method is so flexible to align with various values, the problem with explanation hacking lies in its excessive flexibility. This over-flexibility starts to diminish the credibility and significance of the explanation itself. Modelers can simply reach into their bag of explanations to find anyone that suits your needs.

This problem of CE hacking goes beyond recourse. It is a general CE pitfall that faces all counterfactual explanation methods. And it is a pitfall that seems unique to counterfactual methods.[9] That said, recourse CEs have a *greater risk* of problematic explanation hacking. Given the number of hidden assumptions regarding feasibility and actionability (sect. 3.3.), the way recourse CE can deflect responsibility (sect. 3.2.), and the emphasis on aligning each new CE with new stakeholder values, is the perfect recipe for doing a motivated search through the counterfactual explanation space. Explainers can ignore other counterfactuals that might be a more principal reason for the model decision, such as the feature that holds the most weight for a decision.

## 4. Understanding-first

As an alternative to the recourse-first view for CE, we propose the following understanding-first view.

**Understanding-first:**
Providing users with explanations that aim at understanding AI decisions is (ethically) preferable than recourse explanations.

Understanding is an epistemic state that involves grasping or knowing why something is the case. Understanding is thought to have a high degree of epistemic value (Grimm 2017), can increase our epistemic agency (Kelp 2015), and has social value (Hannon 2018). Moreover, since a principal way of gaining understanding is through knowing an explanation (Khalifa 2017), the social value of *explanation* underscores the social value of understanding. In the context of AI, having an explanation is a central way for end-users to have informed self-advocacy against AI decisions (Vredenburgh 2022). Moreover, the understanding-first view can still contribute to temporally extended agency, like recourse. Understanding the reasons for decisions can help us with our long-term planning even if the information we receive is not by default actionable. Uncovering which possibilities are not open to us can still facilitate temporarily extended agency.[10]

---

[9] Other types of post-hoc explanation methods have been found to be misleading and subject to manipulation (Slack et al. 2020, Dombrowski et al. 2019, Krishna et al. 2022). However, the manipulation of these methods isn't explanation hacking in the sense of actively searching among several possible explanations. Instead, they break by the introduction of noise or by creating adversarial examples. This is a different problem than the one identified here.

In the remainder of this chapter, we introduce important principles that underlie the understanding-first CE methodology, and argue that the understanding-first view does not face the same problems as recourse CE. Thus, we argue that the primary aim and value CE methods should aim towards is understanding-first.

### 4.1. What is an understanding-first CE?

An understanding-first CE places the emphasis on getting a better understanding of the principal-reasons for the model decision. In this chapter, we do not provide concrete guidance for developing understanding-first CE methods. Our aim is to direct future research into constraining CE methods based on a set of understanding-first principles, and to push back on the momentum of recourse CE methods.

The epistemology of understanding is a rich field within epistemology.[11] For the purposes of this chapter, we will not adopt any particular view of understanding and instead focus on two specific features of understanding that are largely shared across the understanding literature: i) someone understands if they are able to answer important why or what-if questions; ii) the content of one's understanding consists of grasping a core set of dependencies, and also more peripheral dependencies about the thing being understood.

First, if someone understands something, they are able to answer important why- or what-if questions about it. For example, for someone to understand why it rains, they are able to answer various questions about the importance that various cloud formations have on rain patterns. They may also be able to answer what-if questions concerning unactualized possibilities, such as how various stages of climate change might impact rain cycles. In the case of AI decisions, counterfactual explanation methods seem especially well-suited for contributing to understanding because CEs provide users with what-if-things-had-been-different scenarios enabling them to answer various what-if questions about the model's behavior.

---

[10] The interplay between epistemology and ethics in AI is complex, see Grote (*forthcoming*).

[11] See Páez (Chapter 9 in this volume) for a nice overview on the epistemology of understanding.

Second, we can separate between the *core* dependencies of why something is the case and more *peripheral* dependencies. For example, if we want to understand why apples fall off trees at a certain stage of ripeness, part of the *full* account of why will depend on the gravitational pull of the moon. However, the pull of the moon is not *central* to understanding why, it is just a small influence that lies outside the most impactful factors. The core dependency for understanding why apples fall off trees at a certain stage of ripeness is primarily related to the ripening process of the apple and Earth's gravity. The ripening process causes changes in the apple, such as the weakening of the stem, making it more likely to detach from the tree. Simultaneously, Earth's gravitational force acts upon the now-loosened apple, causing it to fall.

Some even suggest that if someone's beliefs about peripheral dependencies are all false, they can still have understanding, as long as the core is true (Mizrahi 2012). The specific features that separate the core from the periphery are not straightforward. It is still an active problem in the epistemology of understanding literature. Mizrahi associates the core with grasping a scientific law, but Lawler (2021) takes issue with this restrictive conception. Alternatively, notions of indispensability run large (e.g. Rice 2019). Lawler and Sullivan (2021) provide several intuitive features that capture the core of an *explanation,* such as a central causal mechanism, that there is a non-triviality between certain explanation propositions and the main explanatory power of the explanation, and that the explanatory propositions that uniquely discriminate a particular explanation from some other explanation are part of the core.[12]

An understanding-first CE seeks to uncover the *core* features that give rise to an AI decision from the features that lie on the periphery. How might we separate the *core* set of counterfactuals that explain an AI decision from more peripheral counterfactuals? Baron (2022) argues that CE methods that are embedded within a Pearl-Woodwardean framework of causal interventions are necessary for CE methods to give us true understanding of the system and to establish true causal dependence of decisions (see also Buijsman 2022). On this approach, developing CE methods built on a Pearl-Woodwardean framework could, in principle, give us the core counterfactuals.

For other scientific explanations, Lawler and Sullivan (2020) suggest that the core could be illuminated with robustness analysis, where stable features are the core features (see also Weisberg 2006). In the case of CE methods, aggregating across CEs for different decision subjects could be a way of locating the features that are globally salient for determining the model's decision. Restricting CEs to only include sets of stable global features might be a way of capturing this type of robustness.[13] Closeness to the decision boundary is an existing desiderata in current CE methods that seems like a well-grounded epistemic assumption that gets to the core of an AI decision, but faces its own challenges (see footnote 3). Other desiderata like sparsity might also have epistemic importance for understanding since fewer features might generalize across a large set of instances.[14] Páez (chapter 9 in this volume) suggests that another desiderata important for measuring understanding in the context of CEs is the specific epistemic context. In some contexts, the core causal map might need to be sufficiently detailed, whereas in other contexts a more rough and sparse core is enough for understanding. The understanding-first approach urges CE researchers to develop methods that seek to uncover the core counterfactuals, and that inspiration from philosophy of science and epistemology could be useful.

---

[12] Miller (2019) highlights that good explanations for understanding hinge not on the most powerful causal factors but rather on those most surprising to the recipient.

[13] For a discussion on robustness analysis in the case of ML see Freiesleben and Grote (2023).

One counter-intuitive implication that results from taking this understanding-first perspective, is that it is not necessarily problematic CE methods to allow for impossible real-world counterfactual changes. If the AI model is trained and developed in such a way that it allows for correlating input features in ways that are physically impossible, then it is fair game, if not necessary, to allow for those same impossible correlations to be reflected in the CE of the AI system.[15] For example, if the training process of an AI model allows for decreases in age to develop model correlations and affect model decisions, then we should still allow that assumption into the CE method–since on the understanding-first view–our main goal is understanding the AI system. If, on the other hand, modelers used methods that prevented unrealistic manipulations of data, then these same assumptions must also be built into the CE method. Notice that this is a large departure from recourse CE methods, since for recourse-first the underlying aim is feasible behavior adaptation on the part of the data subject, not understanding the AI decision.

One final note about the understanding-first approach developed here is the distinction between user-understanding and the epistemic norms that can capture core model dependencies. There is a difference between having an explanation that is accurate in the sense of providing all the input features necessary that captures the core of why something is the case, and having an explanation that is legible or intelligible to various people. The latter is something that greatly depends on people's previous familiarity with the subject matter and other psychological aspects. These research streams are separate. Finding the core understanding-first CE is one research stream, where the other research stream is presenting this information to users.[16] It is possible that CE methods are best suited for finding the core dependencies that govern AI decisions, but are not best suited for conveying this information to users (van der Waa et al. 2021).[17] Or vice versa. It might be that other xAI techniques are better suited for finding the core dependencies, but users benefit most if the information is given to them in a counterfactual structure. Often this distinction is collapsed in xAI research, where xAI methods are validated in part by user-studies (Byrne, 2019). User-studies on xAI methods are necessary to make sure users actually understand explanations, but are not a way of locating core model dependencies that give rise to AI decisions.

---

[14] Here work on the unification view of explanation and understanding would be helpful (e.g. Gijsbers 2013). On the unification view, covering more instances increases the amount of understanding.

[15] See Chen et al. (2020) for a similar discussion for SHAP.

[16] See Beisbart (chapter 1 this volume).

[17] See Páez (chapter 9 this volume) for a discussion on user perceptions of CEs.

## 4.2. Responsibility

Understanding-first CEs can alleviate some of the worries related to deflecting responsibility associated with recourse-first CEs. Recall that recourse explanations necessarily, by providing behavioral adaptation guidance to individuals, implicitly place responsibility for the decision on those individuals. However, since current AI systems face issues of model bias and robustness problems, the responsibility for poor AI decisions likely lies elsewhere. Understanding-first CEs need not face this problem. First, CEs that aim at understanding, in principle, would uncover model bias or model unfairness if it was part of the core reasons for a model decision. If an AI system relies on core reasons that underscore bias or unfairness (even through proxies), then data subjects and end-users will have access to this information. For example, in the bank loan case, the retirement counterfactual would be provided as an understanding-first CE because it was the core reason for the decision. Notice that even though the retirement counterfactual is still an individual-based explanation, it shifts the burden of responsibility back on the bank (and the local and state government) for making decisions based on age.

Second, understanding-first CEs also have the flexibility to include socio-structural explanations as a component into the AI decisions. In cases where societal factors are a core contributor to the model decision, then it is even necessary to include these factors in an explanation. While perhaps we can introduce the concept of group recourse, where recourse explanations can go beyond individual actions to collective actions, this can further exacerbate worries of deflecting responsibility onto groups or collectives that may themselves still be marginalized. On the other hand, understanding-first CEs could still point to social structures, but how these structures should be revised would be left open.

## 4.3. Hidden assumptions

In this section, we discuss how the understanding-first view can address some of the ethical pitfalls linked to the hidden assumptions inherent in the recourse-first approach, as previously explored in Section 3.3.

First, the understanding-first perspective offers a considerable advantage over recourse-first explanations, notably in its potential to recognize and integrate systemic and structural factors within counterfactual explanations. Unlike recourse CE, which tends to focus on individual-level changes while often overlooking broader societal contexts, the understanding-first approach is poised to acknowledge and address wider systemic and structural constraints. For instance, a typical recourse-first CE might suggest: "If you had a career in engineering instead of teaching, you would have been approved for the loan." This advice, while direct, disregards systemic issues like employment discrimination, job market saturation, or educational barriers, thereby placing an unfair burden on the individual. In contrast, understanding-first explanations could incorporate socio-structural factors into their framework, highlighting how AI systems make their decisions. For example, that banks have the freedom to make loan decisions based on job-type instead of a collection of other more equitable factors. Of course, this assumes that the AI system possesses extensive background knowledge from socio-structural data analysis. While this is a significant requirement, the potential for such comprehensive explanations certainly exists.

Second, the understanding-first view can circumvent the pitfalls of oversimplified explanations that often arise in recourse-first approaches. By eschewing a one-size-fits-all approach to CE, it treats users as knowledgeable participants. This view assumes respect for the users' ability to understand multiple layers of complex information, aligning more closely with principles of dignity and autonomy in legitimate decision-making. Users who are well-informed about the nuances of AI decision-making are better equipped to adjust their actions and expectations accordingly and engaged in informed self-advocacy.

Third, the understanding-first view facilitates the legitimacy of engaging with contestability. Instead of prioritizing immediate actionability, users are more likely to trust proper CEs when they have a clearer epistemic sense of how AI decisions are made, ensuring that their interaction with AI is based on a solid foundation of knowledge, rather than any systemic or organizational incentives for manipulating the users by prioritizing actionability. The understanding-first view enables users to grasp in a richer way the underlying principles and logic of AI decisions. As compared to the recourse-first view, understanding-first is analogous to teaching someone the rules of a game rather than just what to do for a specific move; the former equips them to strategize better.

Fourth, the concern about normative injections in CE pertains to both recourse-first and understanding-first views. For example, an AI system in a hiring scenario might generate a CE like, "If you had not been Black, your application would have been shortlisted." As Kasirzadeh and Smart (2021) argue, this statement requires deep-seated ontological and epistemological assumptions for its validity. The AI system must have a preconceived definition of race, particularly the construct of being Black. The manner in which race is conceptualized and integrated into the model's space of analytical knowledge significantly influences the outcomes and interpretations of these CEs. However, we believe that the issue of normative injection, particularly regarding actionability and specific individual-level recourse explanations, is less pronounced in understanding-first CEs. By allowing for systemic considerations and the possibility of downplaying the individual attributes such as race or gender, understanding-first CEs can overcome many of the ethical pitfalls associated with imposing normative judgments based on these categories. This approach lends itself to a more holistic and contextually aware form of counterfactual reasoning that acknowledges the complex interplay of societal factors and individual circumstances.

*4.4. Explanation hacking*

Every CE method will have some vulnerability to explanation hacking since the method necessarily involves exploring a larger counterfactual space. However, taking an understanding-first approach reduces this vulnerability. First, the understanding-first approach chiefly concerns finding the core features that give rise to an AI decision. This prevents motivated searching from explanations that already confirm one's beliefs or values. This is important for thinking about the difference between illusory understanding and genuine understanding (e.g. Trout 2022). Illusory understanding comes with the sense of understanding, but without actually grasping the true dependencies underlying a phenomena, or in this case, the reasons for an AI decision. Understanding-first CE methods would be built surrounding what the actual core dependencies are that can make for genuine understanding.

As an extra safeguard, we recommend that in order to prevent explanation hacking, researchers should 'pre-register' the assumptions and values that are embedded in their CE methods, so that the methods can be audited for fairwashing and other biases. This means embracing a checklist like what Kasirzadeh and Smart (2021) suggest, but with awareness of the aims of explanation: recourse or understanding or both. Even in designing understanding-first methods, it is important to first lay a groundwork and motivation for the specific epistemic norms that are central in the method, and what choices are made in the face of mutually conflicting CEs. Taking an understanding-first approach pushes research in a direction that aims to find well-grounded metrics based in epistemically sound principles and developing methods based on those principles. Moreover, we should stop engaging in real time, post-hoc, model exploration that allows users, industry, or developers to justify their model based on flipping through several candidate CEs. While this practice may still be useful for users to find actionable advice or recommendations, it is important to separate the difference between explaining an AI decision versus what action guiding possibilities are open to individuals.

**5. Discussion**

In this chapter, we argued that the trend in xAI research on developing feasible recourse explanations is not the preferable solution for explaining AI decisions. When an AI system makes a decision that affects our lives, we want to know *why* this decision was made. We argued that taking an understanding-first approach to answering this why-question better conforms to other social values concerning responsibility, trust, and informed self-advocacy.

Before concluding, we consider a few objections to the understanding-first view and point to future research directions.

## 5.1. Objections and replies

**Objection 1:** *Since recourse explanations still give you counterfactual inferences, don't they by default also give you understanding? What is the real tension here?*

**Reply 1:** According to minimal accounts of understanding in epistemology *any* answer to a why-question or a what-if question is tantamount to having some small degree of understanding.[18] On this view, recourse explanations provide end-users with understanding, simply in virtue of providing what-if information. Even the recourse explanation in the APARTMENT FIRE case urging tenants to install water accent walls to prevent fires, would still count as providing understanding on these minimal accounts of understanding, since the tenants grasp dependencies that dry walls have on fire strength. However, minimal views of understanding fail to capture what other epistemologists have noted about understanding that there are core dependencies and more peripheral dependencies. If we want attributions of understanding to signal real epistemic success and the holders of *good* explanations (Hannon 2018), what we are really after is that the understander grasps the core dependencies. Similarly, Baron (2022) makes a distinction between basic causal knowledge of an AI system—that could be given by *any* CE—and a complete causal picture that is necessary for understanding why the AI made its decision. Given the problem of explanation-hacking with CE methods, basic causal knowledge of knowing a single remote counterfactual renders the understanding we gain to be borderline trivial. So, if the minimal view of understanding is correct, a simple revision to the understanding-first view--to a *deep* understanding-first view—is necessary. It is not just any CE that gives us deep understanding, only the core CEs give us deep understanding. Recourse explanations that are on the periphery fail to provide deep understanding.

**Objection 2:** *How can you really say that one counterfactual explanation is 'the reason' for a decision, when other counterfactual explanations are strictly speaking also true reasons?*

**Reply 2:** In addressing this objection, it is crucial to distinguish between "a reason" and "core or central reasons" in the context of CEs. A CE, by its nature, posits an alternate scenario that could have led to the same outcome, suggesting that multiple explanations can coexist as valid reasons. However, identifying "the core or central reasons" typically involves prioritizing one explanation over others based on its relative significance or causal weight in the specific context. In any decision-making scenario, numerous factors contribute to the final outcome, each representing a potential CE. The challenge lies in evaluating these factors to determine which holds the most weight. This evaluation often involves an analysis of the directness of the cause, its historical precedence, or its explanatory power in simplifying or understanding of the event. So, when assessing that one CE is "the core or central reason" for a decision, it implies that this reason is identified as the most significant or influential among several true reasons. It does not negate the validity of other explanations but recognizes a hierarchy of influences, where the primary reason has a most direct or substantial impact on the outcome than others. So, while multiple CEs can be true, assessing one as "the core or central reason" involves a reasoned judgment about its relative importance in the specific context of the decision. Understanding-first principles urges researchers to develop methods that are able to make such judgments on sound epistemic principles. Importantly, it might very well be that CE methods are ill suited for the task of determining causal weight and isolating core reasons from peripheral reasons. If so, understanding-first recommends abandoning CE methods for those that are more suited for providing understanding.

---

[18] See Hubert and Malfatti (2022) for a nice discussion on the degrees and threshold for understanding.

**Objection 3:** *If people want actionable explanations, give the people what they want!*

**Reply 3:** We are not advocating against giving feasible recourse CEs full stop. Nor are we advocating against doing research on developing methods for extracting feasible recourse CEs. We are arguing for three central claims. First, feasible recourse CEs are not the gold ethical standard of CEs. Second, we have an obligation to always include an understanding-first CE with AI decisions. In some cases, the understanding-first CE and the recourse CE will overlap, in cases where there isn't overlap it is possible to provide both a recourse CE and an understanding CE. However, in this latter case we suggest, following Sullivan and Verreault-Julien (2022) in using the language of recourse *recommendations*, instead of recourse *explanations*.[19] Third, we are highlighting the urgency to move away from real time model exploration of the counterfactual space that some researchers are purposely building into explanation interfaces (e.g. Bove et al. 2023, Cheng et al. 2020, and Gomez et al. 2020). Instead, we urge CE researchers to 'pre-register' their assumptions and calibrated methods for generating CE, and then stick to these pre-registered methods, even if it means giving an end-user a CE that doesn't align with their values. Doing so would signal to the end-user that maybe the system itself does not align with their values, providing them with genuine grounds to trust or distrust the system.

---

[19] See also Baron (2022), who argues that we should provide both advice and complete causal information for why the decision was reached.

## 5.2. Future research directions

In this chapter, we focused exclusively on counterfactual explanation methods. However, recourse explanations do not need to be counterfactual explanations. While counterfactual explanations are a common method used in providing recourse in AI systems, they represent just one approach among several possible methods. Other methods may include, for example, non-counterfactual causal reasoning, where the explanation focuses on the causal relationships within the data, or rule-based explanations, where certain rules or conditions are provided for achieving a different outcome. Our arguments may also apply to these other recourse explanations, as long as they are not part of the core reasons for an AI decision.

Lastly, in this chapter, we do not solve the hard problem of how we could go about developing understanding-first CEs. If anything, some of the main problems with CE methods still loom large even on an understanding-first view. CE methods still make several hidden assumptions regarding what kind of features lend themselves to counterfactual reasoning, which are likely misguided, like in the case of sensitive attributes like race and gender. CE methods are still susceptible to explanation-hacking even if we focus on developing understanding explanations, since there can still be motivated searching of the counterfactual space even considering epistemic norms. Lastly, there are hard conceptual and technical challenges to developing CE methods that actually achieve the aim of understanding. It may very well be that given all these considerations that we should opt for a different type of explanation method altogether. We suspect the right solution will involve utilizing a variety of explanation methods (Freiesleben and König 2023), where CE methods are just one aspect to explain why a certain AI decision was reached.

## Acknowledgments

We presented this work at various stages in development at Oxford University, Utrecht University, and the 2023 ESDiT away days. We thank the attendees for their helpful comments and suggestions. We would also like to thank Juan Durán, Timo Freiesleben, Thomas Grote, Andrés Páez, and Yeji Streppel for their comments and suggestions. This work is supported by the Netherlands Organization for Scientific Research (NWO grant number VI.Veni.201F.051) and the research programme Ethics of Socially Disruptive Technologies funded by the Gravitation programme of the Dutch Ministry of Education, Culture, and Science and the Netherlands Organization for Scientific Research (NWO grant number 024.004.031).